\documentclass[10pt,preprint,a4paper]{aastex}

\usepackage{amsmath}                
\usepackage{amsfonts}               
\usepackage{amssymb}                
\usepackage{epsfig}                 

\newcommand{\cm}{{~\rm cm}}
\newcommand{\s}{{~\rm s}}
\newcommand{\km}{{~\rm km}}
\newcommand{\g}{{~\rm g}}
\newcommand{\K}{{~\rm K}}
\newcommand{\erg}{{~\rm erg}}
\newcommand{\yr}{{~\rm yr}}

\newcommand{\days}{{~\rm days}}

\begin{document}

\title{EXPLAINING THE SUPERNOVA IMPOSTOR SN 2009ip AS MERGERBURST}

\author{Noam Soker\altaffilmark{1}, Amit Kashi\altaffilmark{2}}

\altaffiltext{1}{Department of Physics, Technion -- Israel Institute of Technology, Haifa 32000, Israel; soker@physics.technion.ac.il}
\altaffiltext{2}{Department of Physics and Astronomy, University of Nevada, Las Vegas, 4505 S. Maryland Pkwy, Las Vegas, NV, 89154-4002, USA; kashia@physics.unlv.edu}

\begin{abstract}
We propose that the energetic major outburst of the supernova (SN) impostor SN 2009ip in September 2012 (outburst 2012b) was a \emph{mergerburst} event,
where two massive stars merged.
The previous outbursts of 2009 and 2011 might have occurred near periastron passages of the binary system prior to the merger,
in a similar manner to the luminosity peaks in the nineteenth century Great Eruption of the massive binary system Eta Carinae.
The major 2012b outburst and the 2012a pre-outburst, resemble the light curve of the mergerburst event V838~Mon.
A merger of an evolved star with a mass of $M_1 \sim 60$--$100 M_\odot$ and a secondary main sequence star of $M_2 \sim 0.2$--$0.5 M_1$ can account for the energy of SN~2009ip
and for the high velocities of the ejected gas.
The ejected nebula is expected to have a non-spherical structure, e.g. bipolar or even a more complicated morphology.
\end{abstract}


\section{INTRODUCTION}
\label{sec:introduction}

SN~2009ip is an enigmatic luminous blue variable (LBV) star \citep{Berger2009}, located in the spiral galaxy NGC~7259.
Since its first observed outburst in 2009 \citep{Maza2009} it had suffered two more long outbursts (e.g., \citealt{Drake2012})
of 3--4 magnitudes brightening in the V band, in September 2011 and August 2012 (outburst 2012a),
and an outburst of $\sim 7$ magnitudes in September 2012 (outburst 2012b).
There was a shorter and less pronounced outburst in 2012 (e.g. \citealt{Mauerhan2012}).
The peak bolometric luminosity of the 2012b outburst was $L_{\rm p} = 8 \times 10^{42} \erg \s^{-1}$ \citep{Pastorello2012}.
Retroactive searches revealed the progenitor of SN~2009ip in observations done several years earlier \citep{Miller2009}.
A detailed summary of the SN~2009ip evolution is given by \cite{Levesque2012}.

The peak luminosity of outburst 2012b, and the P-Cygni absorption wings that extend to $\sim 13\,000 \km \s^{-1}$ brought \cite{Mauerhan2012}
to suggest that the 2012 outbursts were a result of a supernovae (SN) explosion. They suggest that the first 2012a outburst was the SN event,
while the major 2012b outburst is a result of the collision of the SN ejecta with previously ejected gas.
\cite{Pastorello2012} noted that the September 2011 outburst already had such wide P-Cygni wings, and raised the possibility
that the 2012b outburst was not a SN after all.
In that respect \cite{Smithetal2010, Smithetal2011} and \cite{Mauerhan2012} (also \citealt{Levesque2012}) already made a connection between SN~2009ip and $\eta$ Carinae ($\eta$ Car),
which also had a small fraction of its ejected mass from its 1837--1856 Great Eruption (GE) moving at high velocities \citep{Weisetal2001} of
up to $\sim 5000 \km \s^{-1}$ \citep{Smith2008}.

The high luminosity of the 2012b outburst was attributed to the collision of fast ejecta with a slower shell \citep{Mauerhan2012, Prieto2012}.
The brightening was observed in other bands beside the visible one.
\cite{Marguttietal2012} reported that the re-brightening in the UV is more extreme than that in V.
A transient X-ray source with $L_x = 2.3 \pm 0.5 \times 10^{39} \erg \s^{-1}$ that may be associated with the same outburst has been reported by \cite{MarguttiSoder2012}.

\cite{Foley2011} and \cite{Smithetal2010} determined the progenitor star of SN~2009ip to have $M > 60 M_\odot$, and found its 2009 outburst
to be similar to LBV giant outbursts, such as the GE of $\eta$~Car (for more on the GE see, e.g., \citealt{DavidsonHumphreys1997}).
As the sharp peaks in the GE of $\eta$~Car were triggered by binary interaction at periastron passages
\citep{Damineli1996, KashiSoker2010a, SmithFrew2011}, it might be that the same had occurred for SN~2009ip, with the
last energetic outburst being an extreme case of binary interaction, namely a merger event.
This is discussed in section \ref{sec:merger}.
In a recent paper \cite{Levesque2012} discuss the presence of a binary companion, but not periastron interaction.
Our short summary is in section \ref{sec:summary}.

\section{A MERGERBURST MODEL}
\label{sec:merger}

\begin{figure}[!t]
\center
\resizebox{0.89\textwidth}{!}{\includegraphics{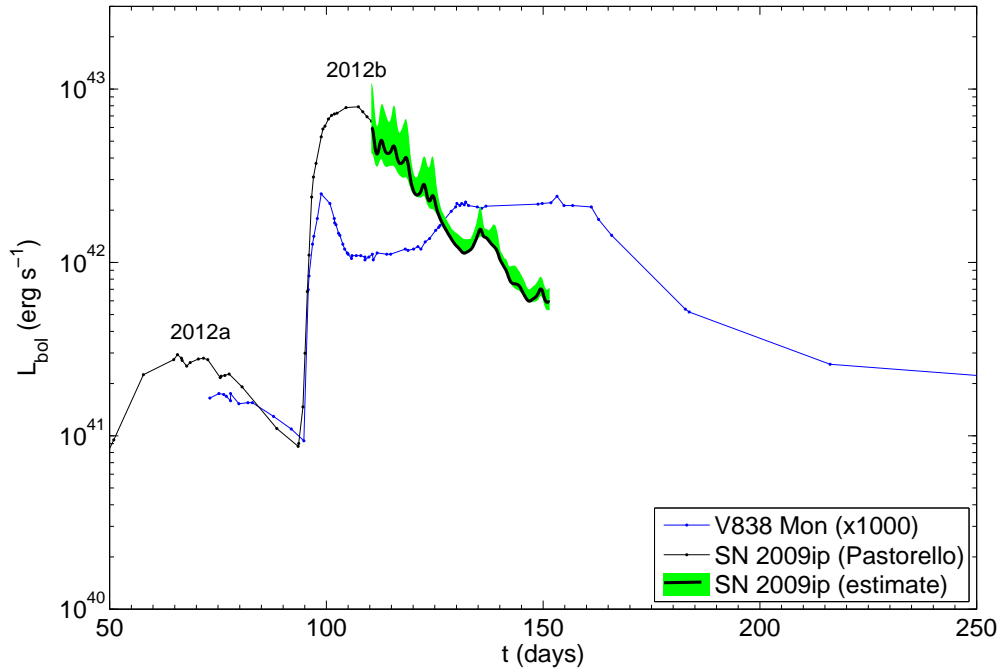}}
\caption{Comparison of the light curves of V838~Mon in 2002 and SN~2009ip in 2012.
The time for SN~2009ip is given in days after JD 2456100 (days 50, 100 and 150 correspond to August 10, September 29, and November 18, respectively).
The light curve of V838~Mon is taken from \cite{Tylenda2005} and is multiplied by a factor of 1000.
The light curve of SN~2009ip to day 112 in the graph is from \cite{Pastorello2012}. Beyond that time the light curve is an estimate
we have made based on the V and I photometry taken from the UIS Henry Barber Research Observatory.
The error of our estimate appears as a green shade.
It is evident that the large peaks are not just minor fluctuations.
}
\label{fig:lightcurves1}
\end{figure}

\cite{Pastorello2012} give the {{{{ bolometric }}}} light curve of SN~2009ip up to few days after the peak.
To estimate the {{{ bolometric}}} luminosity and the effective temperature beyond that,
we use the V and V-I photometry from the open access site of the
UIS Henry Barber Research Observatory\footnote[1]{\url{https://edocs.uis.edu/jmart5/www/barber/SN2009ip.html}},
(majority of data are from \citealt{Hambsch2012}; see \citealt{Martinetal2012} and ATel \#4539).
We use the bolometric correction from \cite{{BerstenHamuy2009}} to calculate the bolometric luminosity out of the V$-$I color.
{{{{ This correction was developed for Type IIP SNe that have different spectra and evolution than SN 2009ip.
We take this into account as described next.
The systematic error in the bolometric correction (BC) comes from the absence of a detailed spectra that does not allow us a detailed BC calculation.
Comparing our derived bolometric light curve to the accurate bolometric light curve of \cite{Pastorello2012}
in times both data sets are available (from 4 days before the peak up to 10 days after), we find  that we underestimate the bolometric luminosity by a factor of 1.7.
We therefore correct our estimated light-curve by this factor.
 The random errors in our estimated light curve are calculated from the error estimates of the V$-$I observations.}}}}
As we do not have the spectral evolution, our luminosity and radius are crude, and only serve to estimate the total radiated energy.
The light curve from \cite{Pastorello2012}, with our luminosity estimation beyond their plot (112 day in the graph) is presented
in Figure \ref{fig:lightcurves1}, together with the light curve of V838~Mon taken from \cite{Tylenda2005}.
The luminosity of V838~Mon is multiplied by $1000$.

V838~Mon and SN~2009ip show the following similarities:
($i$) A pre-main outburst that lasted for about a month.
($ii$) A sharp rise towards the main outburst. The luminosity increase by a factor of 10 within few days.
($iii$) The decline is not monotonic. {{{{ Both objects show another peak about a month after the main peak}}}}.
V838~Mon regains its peak luminosity, while SN~2009ip suffers minor luminosity peaks.
Recent evolution of the V photometry (two weeks beyond our last point) shows that the behavior continues to resemble that of V838~Mon in decline,
but we have no photometry from another filter to estimate the bolometric luminosity.

Although the general light curves are qualitatively similar, we do not expect the two light curves to be exactly the same.
The reason is that the light curve is determined in part by interaction of ejected shells, which are ejected at different times
and velocities for each merger event.
\cite{Tylenda2005} identified 3 shells that were ejected within a few days during the merger process of V838~Mon.
It is possible that there are 2 and more fast shells that were ejected during the merger of SN~2009ip.
The second shell was of low mass and is responsible for the small peak in the light curve of SN~2009ip at $t \sim 136 \days$ in Figure \ref{fig:lightcurves1}.
Interestingly, during this small peak, on November 6, 2012 ($138 \days$ in the graph), \cite{Mauerhan2012} reported the reappearance of
Broad H~I P-Cygni profiles, with FWHM$\sim 5000$--$6000 \km \s^{-1}$ for the main emission profile and a blue absorption edge out to $\sim 15\,000 \km \s^{-1}$.

In V838~Mon the merger is thought to have occurred at the beginning of the main outburst.
In the case of SN~2009ip the view is that the main ejection occurred during the 2012a outburst, before the main peak (2012b); the
main peak itself is due to the collision of the 2012a ejecta with previously ejected gas.
If a shell is ejected at the beginning of the 2012b major outburst of SN~2009ip (as in V838~Mon),
then it reaches the maximum radius of $\sim 2 \times 10^{15} \cm$ within
20 days from the beginning of outburst 2012b.
Before the 2012b outburst the {{{{ photospheric }}}} radius was at $\sim 10^{15} \cm$.
{{{{ Namely, the 2012a outburst photosphere was at $\sim 10^{15} \cm$. The 2012a outburst did not have much kinetic energy,
and it lost a large part of it to radiation. When the 2012b outburst occurred, it catched-up with this and previously ejected gas and turned lots of
kinetic energy (not all of it) to radiation. The bright 2012b outburst was observed when its gas broke out of the
photosphere from the 2012a outburst. This is what happened with the `three shells' in V838 Mon as presented by \cite{Tylenda2005}. }}}}

This expansion corresponds to an average  velocity of $v_{\rm shell} \simeq 6000 \km \s^{-1}$, with a maximum photospheric expansion rate
during the rise of $\sim 10^4 \km \s^{-1}$.
The photospheric temperature is $>6000 \K$, so the opacity at these low densities is $\kappa \ga 0.1 \cm^{2} \g^{-1}$.
A shell of mass $M_{\rm shell} \sim 0.15 M_\odot$ expanding at $\sim 6000 \km \s^{-1}$ can account for the photosphere at $R \simeq 2 \times 10^{15} \cm$.
Such a shell has a kinetic energy of $E_{\rm shell} \sim 5 \times 10^{49} \erg$.
{{{{ The mass of $0.15 M_\odot$ is set as a minimum to account for the photosphere and an energy of $5\times 10^{49} \erg$ at a velocity of $6000 \km \s^{-1}$.
The energy that is available is larger (see below), so a mass of up to $\sim 0.5 M_\odot$ can be ejected at $6000 \km \s^{-1}$
(with small amounts at higher velocities), and more mass can flow at lower velocities. Only a fraction of the few$\times M_\odot$ that are strongly perturbed in
the primary envelope by the merger process are escaping at very high velocities. Most of the mass expands at slower speeds and a large fraction of the mass
inflates an envelope and not even escape the system, if the process is similar to that proposed for V838~Mon by \cite{SokerTylenda2006}.  }}}}

Although we expect a bipolar outflow, or even a more complicated flow shaped by the merger process, in this simple calculation we assumed a spherical geometry.
A directional flow requires less energy, and can account for high velocities if the fast flow points at the observer.

The obtained SN~2009ip total radiated energy is $E_{\rm rad} \simeq 2 \times 10^{49} \erg$. 
The fast ejecta from the 2012 outburst collided with the
previously ejected gas, and a large fraction of the kinetic energy was transferred to radiation.
This is not as in V838~Mon, where only a small fraction, $\sim 10 \%$, of the released energy was radiated.
We crudely estimate the total energy (kinetic + radiated) of SN~2009ip over the several years variability to be $E_{\rm ip} \simeq 5 \times 10^{49} \erg$.
This is $\sim 50$--$150$ times the energy in V838~Mon, where the uncertainty is in the kinetic energy of V838~Mon.
The model of V838~Mon assumes the merging of a $\sim 0.3$--$0.5 M_\odot$ per-main sequence star with a $\sim 6 M_\odot$ main sequence star \citep{TylendaSoker2006}.

For our mergerburst model of SN~2009ip we consider two massive stars on a very eccentric orbit with a periastron distance of
$r_p \simeq 0.2$--$0.5 R_1 << a $, where $R_1$ is the primary radius and $a$ is the semi-major axis.
We take one evolved star of $(M_1,R_1)=(80 M_\odot, 40 R_\odot)$, and another main sequence star of $M_2=20 M_\odot$.
We consider a situation where the final merger (September 2012) occurred when the secondary entered the primary outer envelope, with a periastron
distance of $r_p \simeq 10 R_\odot \simeq 0.25 R_1$.
The secondary explosively {{{{ inflated and }}}}  ejected the outer envelope of the primary.
The energy available in the merger process is
\begin{equation}
    E_{\rm mer} \simeq  \frac{1}{2} \frac{G M_1 M_2}{r_p} = 3 \times 10^{50}
    \frac{M_1 M_2}{(40 M_\odot)^2}
    \left( \frac{r_p}{10 R_\odot} \right)^{-1}  \erg.
\label{eq:Emer1}
\end{equation}
We did not take into account the energy that is needed to unbind the secondary star, {{{ as this process will take longer.
More than that, the mass originated from the secondary will sink inside the primary, releasing more energy. }}}
The conclusion is that a massive stellar merger can supply the required energy.
A large fraction of this energy is channelled to inflate an extended envelope, as in the model for V838~Mon \citep{SokerTylenda2006}.

In the last few periastron passages it is likely that mass was transferred from the primary envelope to the secondary.
Therefore, accretion onto the secondary could have contributed to the luminosity of the outbursts prior to the main 2012b outburst.
We note that \cite{Levesque2012} mention mass transfer to the companion as
a way to form the thin disk in their model for SN~2009ip, but do not consider energy release from this process.
The typical radiated energy in the 2011 peak, for example, is $E_{2011} \simeq 2 \times 10^{47} \erg$. 
However, the kinetic energy of the ejected mass could have been larger.
An amount of energy of $\sim 5 \times 10^{47} \erg$, for example, could have been released by a mass of $0.1 M_\odot$ that was accreted onto a
MS star of $(M_2,R_2)=(20 M_\odot, 6 R_\odot)$, assuming half the released energy was radiated.
{{{{ This point is not directly related to the mergerburst model. We only want to stress that mass transfer in
the strongly interacting binary system can account for the outbursts that preceded the 2012b major outburst. }}}}
In our model, it is likely that part of the mass that was transferred to the secondary star is ejected in a bipolar outflow,
as in the model for the GE of $\eta$ Car that formed the Homunculus \citep{KashiSoker2010a}.

{{{{ The behavior of $\eta$ Car during its Great Eruption (GE) teaches us that not every emission-peak signals a periastron passage.
In Eta Carinae we know the orbital period during the GE was ~5.5 years.
The sharp peaks in the light curve during the GE of $\eta$ Car follow the $\sim 5.5 \yr$ orbital period at that time
\citep{KashiSoker2010a}. In addition there are emission peaks at shorter time scales, most notably in 1845 \citep{SmithFrew2011},
just two years after the previous peak, when the system was close to its apastron.
Such emission peaks can result from variations in mass transfer rates that can result not only from periastron passages,
but also from stochastic mass loss from the unstable primary star.}}}}

Most challenging to the mergerburst model is the high velocity observed. Preliminary results from 3D merger simulations of our
group (Tsebrenko et al., in preparation) show that a non-negligible mass is expelled at velocities
of up to $\xi \simeq 3$--$5$ times the merging relative velocity of the two stars. The high velocity outflow is directional in a wide angle.
Scaling with the above parameters we find
\begin{equation}
    v_{\rm max}  =  \xi \left[ \frac{2 G (M_1 + M_2)}{r_p} \right]^{1/2} = 7\,800
    \left( \frac{\xi}{4} \right)
   \left( \frac{M_1 +M_2}{100 M_\odot} \right)^{1/2}
   \left( \frac{r_p}{10 R_\odot} \right)^{-1/2}  \km \s^{-1}.
\label{eq:Velocity}
\end{equation}
We note that the amount of mass outside a radius of $10 R_\odot$ ($5 R_\odot$) in our model of $(M_1,R_1)=(80M_\odot, 40 R_\odot)$ is $3 M_\odot$ ($10 M_\odot$).
Therefore, the secondary of $\sim 20 M_\odot$ can dive to a distance of $\sim 5-10 R_\odot$ quite easily.
The results of the merger 3D simulations will be presented in a forthcoming paper.

The main conclusion here, based on the light curve similarity to that of V838~Mon and the energetic considerations,
is that the outbursts of SN~2009ip could be accounted for by interaction of a highly eccentric binary system.
The last most energetic outburst (2012b) was due to the final merger of the two stars.

\section{SUMMARY}
\label{sec:summary}

There is a growing number of transient systems with outbursts that can be attributed to main sequence and evolved stars
that accrete mass at very high rates \citep{Kashisoker2010b}.
An extreme case of mass transfer is the merger of two stars, a process named \emph{mergerburst}.
Most prominent are V838~Monocerotis \citep{TylendaSoker2006} and V1309~Scorpii \citep{Tylendetal2011a}.
Events of high mass transfer rates in LBV massive binary stars are though to have occurred in the nineteenth century
Great and Lesser Eruptions of $\eta$ Car \citep{KashiSoker2010a},
and in the seventeenth century eruption of P~Cygni \citep{Kashi2010}.
In this paper we suggest that SN~2009ip was a massive binary system on a very eccentric orbit. Several periastron passages
caused several outbursts, that ended in a major outburst (2012b) caused by the merger of the two stars.

Support to this suggested mergerburst model come from the similar light curve to that of V838~Mon,
as presented in Figure \ref{fig:lightcurves1}, and energetic arguments presented in section \ref{sec:merger}.
Although luminous SN can have luminosity peaks along their declines (e.g., \citealt{Gal-Yam2012}),
the luminosity of the major outburst of SN~2009ip is like that of a common SN and not like luminous ones.
As well, there is no {{{{ immediate }}}} pre-outburst in common SN.
{{{{ This model needs further study, as SN 2006jc, for example, had a pre-explosion outburst.
However, we note the following.
$(i)$ There are no peaks in the declining light-curve of SN 2006jc as large as the two largest peaks in the light-curve of SN 2009ip.
$(ii)$  In SN 2006jc the pre-outburst is not adjacent to the main outburst, but rather occurred 2 years earlier than than the main peak.
So there could have been a binary interaction, as \cite{Pastorelloetal2007} suggest, but the main outburst was not a merger
similar to that proposed for V838 Mon.
$(iii)$  In 2006jc the broad velocity component can be observed for more than 2 months after peak emission \citep{Pastorelloetal2008}.
In SN 2009ip the narrow lines dominate around the peak; the broad lines, if exist, are much weaker.
Over all, the similarity between 2006jc and 2009ip is inconclusive. }}}}

As for the energetics, the kinetic energy of the ejecta should be determined in the coming months.
The emission and absorption from very fast gas disappeared as of October 2012 \citep{Vinko2012},
(or appeared very weakly; \citealt{Jha2012, Childress2012}) and reappeared during the small peak in the first week of
November (\citealt{Mauerhan2012}; around day 136 in the Fig. \ref{fig:lightcurves1}).
This implies that the kinetic energy of the ejecta is not large. However, if indications for massive fast ejecta reappear such that the kinetic energy is
$E_{\rm{kin}} \ga 2 \times 10^{50} \erg$, then the mergerburst model is pretty much ruled out.

Let us end by listing our predictions for the future evolution of SN~2009ip.
\begin{enumerate}
\item The kinetic energy of the ejecta be $E_{\rm{kin}} < 2 \times 10^{50} \erg$, and more likely $E_{\rm{kin}} \sim 5 \times 10^{49} \erg$.
\item As this is not a SN of a massive star, the total ejected mass is smaller than the mass of the star, and we expect $M_{\rm ej} \sim 1$--$3 M_\odot$.
{{{{ Only a small fraction of this mass moves at the very high velocities of $v \ga 5000 \km \s^{-1}$. }}}}
\item After the ejecta becomes optically thin, a giant star will emerge as the merger product. Whereas the remnant of the V838~Mon is a red giant,
e.g., a $\sim$~M6 red giant after seven years \citep{Tylendaetal2011b}, the much more massive remnant of SN~2009ip is expected to be a hotter giant.
\item The ejecta has a bipolar, or even a more complicated, structure,
similar to the Homunculus--the bipolar nebula of $\eta$ Car. The same prediction is made by \cite{Levesque2012}.
This structure can reveal itself in polarimetry.
\item There will be no more large outbursts (beside possibly in the next several weeks, as in V838~Mon).
Small fluctuations are possible due to the complicated structure of the ejecta.
\end{enumerate}

{{{{ We thank an anonymous referee for very helpful comments. }}}}
This research was supported by the Asher Fund for Space Research at the Technion, and the US-Israel Binational Science Foundation.

{}

\end{document}